\documentclass[twocolumn]{IEEEtran}

\IEEEoverridecommandlockouts
\usepackage{cite}
\usepackage{amsmath,amssymb,amsfonts}
\usepackage{algorithmic}
\usepackage{graphicx}
\usepackage{textcomp}
\usepackage{xcolor}
\def\BibTeX{{\rm B\kern-.05em{\sc i\kern-.025em b}\kern-.08em
    T\kern-.1667em\lower.7ex\hbox{E}\kern-.125emX}}
\begin{document}

\title{High-performance Power Allocation Strategies for Secure Spatial Modulation\\
}

\author{Feng Shu,~Xiaoyu Liu,~Guiyang Xia,~Tingzhen Xu,~Jun~Li,~and~Jiangzhou Wang,~\IEEEmembership{Fellow,~IEEE}

\thanks{This work was supported in part by the National Natural Science Foundation of China (Nos. 61771244, 61472190, 61501238, 61801453, 61702258, and 61271230)).}
\thanks{Feng Shu,~Xiaoyu Liu,~Guiyang Xia,~Tingzhen Xu,~and~Jun~Li are with the School of Electronic and Optical Engineering, Nanjing University of Science and Technology, 210094, CHINA. (Email: shufeng0101 @163.com). }
\thanks{Feng Shu is also with the School of Computer and  information at Fujian Agriculture and Forestry University, Fuzhou, 350002,  China.}
\thanks{Jiangzhou Wang is with the School of Engineering and Digital Arts, University of Kent, Canterbury CT2 7NT, U.K. Email: \{j.z.wang\}@kent.ac.uk.}

}

\maketitle

\begin{abstract}
Optimal power allocation (PA) strategies can make a significant rate improvement in secure spatial modulation (SM). Due to the lack of secrecy rate (SR) closed-form expression in secure SM networks, it is hard to optimize the PA factor. In this paper,  two PA strategies are proposed: gradient descent, and maximum product of signal-to-interference-plus-noise ratio (SINR) and artificial-noise-to-signal-plus-noise ratio (ANSNR)(Max-P-SINR-ANSNR). The former is an iterative method and the latter is a closed-form solution. Compared to the former, the latter is of low-complexity. Simulation results show that the proposed two PA methods can approximately achieve the same SR performance as exhaustive search method and perform far better than three fixed PA ones. With extremely low complexity, the SR performance of the proposed Max-P-SINR-ANSNR  performs slightly better and worse than that of the proposed GD in the low to medium, and high signal-to-noise ratio regions, respectively.
\end{abstract}


\section{Introduction}
In multiple-input-multiple-output (MIMO) systems, spatial modulation (SM) \cite{Mesleh2008Spatial} was proposed as the third method to strike a good balance between spatial multiplexing and diversities while Bell Laboratories Layer Space-Time (BLAST) in \cite{Foschini2010Layered} and space time coding (STC) in \cite{Yu2015Power} were the first two ways. Unlike BLAST and STC, SM exploits both indices of activated antenna and modulation symbols to transmit information, which can increase the spectral efficiency and reduce the complexity and cost of multiple-antenna schemes without deteriorating the end-to-end system performance and still guaranteeing good data rates \cite{Survey2014}.  Compared to BLAST and STC, SM has a higher energy-efficiency due to the use of less active RF chains.

How to enable SM to transmit confidential messages securely is an attractive and significantly important problem \cite{Liang2007Secure, Shu2017Secure,Shu2018two}. In \cite{Wang2015Secrecy}, the authors analyzed the secrecy rate (SR) of SM for multiple-antenna destination and eavesdropper receivers. Instead of typical requirements for eavesdropper channel information, they investigated the security performance through joint signal and interference transmissions. Furthermore, \cite{Liu2017Secure} proposed and investigated a full-duplex receiver assisted secure spatial modulation scheme. It enhances the security performance through the interference sent by the full duplex legitimate receiver. In \cite{Shu2018two}, the authors proposed two novel transmit antenna selection  methods: leakage and maximum SR,  and one generalized Euclidean distance-optimized antenna selection method  for secure SM networks.

In a secure directional modulation system \cite{Wan2018Power},   power allocation (PA) between confidential message and artificial noise (AN) was shown to have an about 60 percent improvement on SR performance. Similarly, PA is also crucial for secure SM with the aid of AN.  In \cite{Wu2015Secret}, the optimal PA factor between signal and interference transmission was given by exhaustive search (ES) for precoding-aided spatial modulation. However, the computational complexity of ES is very high for a very small search step-size. Therefore, a low-complexity PA method is preferred for practical applications. By focusing  on PA  strategies in secure SM, our main contributions in this paper are as follows:
\begin{enumerate}
 \item  We derive an approximate SR expression to the actual SR. Using this approximation, we establish the optimization problem of maximizing SR over PA factor given AN projection matrix. A gradient descent (GD) algorithm is adopted to address this problem. The proposed GD converges to the locally optimal point. However, it is not guaranteed to converge the globally optimal point and may approach the optimal point by increasing the number of random initializations. Additionally, it is also an iterative method, and depend heavily on its termination condition.

 \item  To address the above iterative convergence problem of the proposed GD, a novel method, called maximizing the product of signal-to-interference-plus-noise ratio (SINR) and artificial-noise-to-signal-plus-noise ratio (ISNR)(Max-P-SINR-ANSNR), is proposed to provide a closed-form expression. This significantly reduces the complexity of GD. Simulation results  show that the proposed Max-P-SINR-ANSNR can achieve a SR performance close to that of optimal ES. This makes it become a promising practical PA strategy.
     \end{enumerate}

 The remainder is organized as follows.  Section II describes system model of secure SM system and express the average SR. In Section III,  two  PA strategies are proposed for secure SM. We present our simulation results in Section IV. Finally, we draw conclusions in Section V.

\section{System Model}

\begin{figure}[h]
\centerline{\includegraphics[width=0.36\textwidth]{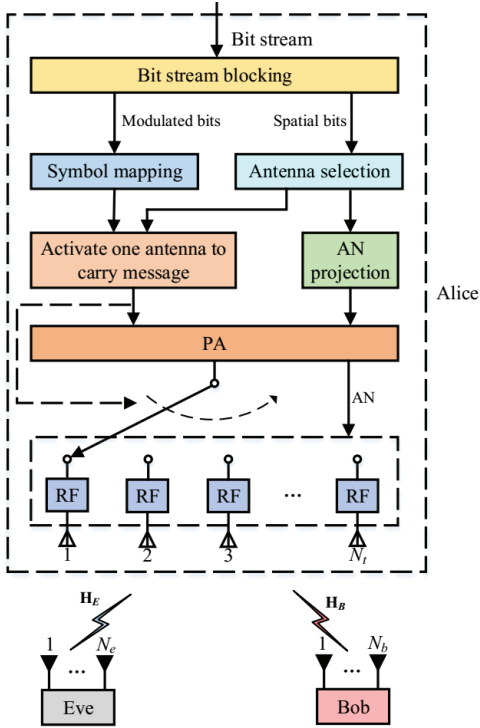}}
\caption{Block diagram of secure SM.}
\label{fig2}
\end{figure}

Fig.~1 sketches a secure SM system with $N_t$ transmit antennas (TAs) at transmitter (Alice).  $N_r$ and $N_e$ receive antennas (RAs) are employed at desired receiver (Bob) and eavesdropping receiver (Eve), respectively. Alice's confidential information sent to Bob from the channel will be intercepted by Eve. Additionally, the size of signal constellation $\mathcal{M}$ is  $M$. As a result, $\textrm{log}_2N_tM$ bits can be transmitted per channel use, where $\textrm{log}_2N_t$ bits are used to select one active antenna and the remaining $\textrm{log}_2M$ bits are used to form a constellation symbol. Similar to the secure SM model in \cite{Shu2018two}, the transmit signal with the help of AN is represented by
\begin{align} \label{x}
\textbf{s}=\sqrt{\beta P}\textbf{e}_ib_j+ \sqrt{(1-\beta)P}\textbf{T}_{AN}\textbf{n},
\end{align}
where $\textbf{x}=\textbf{e}_ib_j$ , $\beta\in[0,1] $ is the PA factor, and $P$ denotes the total transmit power constraint. Here, $\textbf{e}_i$ is the $i$th column of identity matrix $\textbf{I}_{N_t}$, implying that the $i$th antenna is chosen to transmit symbol $b_j$, and $b_j$ is the input symbol equiprobably drawn from a $M$-ary  constellation. In addition, $\textbf{n} \in \mathbb{C}^{N_t \times 1}$ is the AN vector. The receive signals  at the desired and eavesdropping receivers are
\begin{align} \label{y_B}
&\textbf{y}_g=\sqrt{\beta P}\textbf{H}_g\textbf{e}_ib_j+
\sqrt{(1-\beta)P}\textbf{H}_g\textbf{T}_{AN}\textbf{n}+\textbf{n}_g
\end{align}
where $g$ stands for B (Bob) or E(Eve), $\textbf{H}_B$,  and $\textbf{H}_E \in \mathbb{C}^{N_e \times N_t}$ are the channel gain matrices from Alice to Bob and to Eve, with each elements of $\textbf{H}_B$ and $\textbf{H}_E$ obeying the Gaussian distribution with zero mean and unit variance. Additionally, $\textbf{n}_B \in \mathbb{C} ^{N_r \times 1}$ and $\textbf{n}_E \in \mathbb{C} ^{N_e \times 1}$ are the complex Gaussian noises at Bob and Eve with $\textbf{n}_B \sim \mathcal{CN}(0,\sigma_B^2\textbf{I}_{N_r})$ and $\textbf{n}_E \sim \mathcal{CN}(0,\sigma_E^2\textbf{I}_{N_e})$, respectively. Given a specific channel realization,  the mutual information of Bob and Eve  are as follows
\begin{align} \label{I_B} \nonumber
I_B(\textbf{x};\textbf{y}_B')&=\textrm{log}_2N_tM - N_t^{-1}M^{-1}\times \\ & \sum \limits_{i = 1}^{N_tM} {\mathbb{E}_{\textbf{n}_B} \left\{ \textrm{log}_2 \sum \limits_{j=1} ^{N_tM} \textrm{exp} \left(-f_{b,i,j}+\|\textbf{n}_B'\|^2 \right) \right\}}
\end{align}
\begin{align}
\label{I_E} \nonumber
I_E(\textbf{x};\textbf{y}_E')&=\textrm{log}_2N_tM - N_t^{-1}M^{-1}\times \\ & \sum \limits_{m  = 1}^{N_tM} {\mathbb{E}_{\textbf{n}_E'} \left\{ \textrm{log}_2 \sum \limits_{k=1} ^{N_tM} \textrm{exp} \left( -f_{e,m,k}+\|\textbf{n}_E'\|^2 \right) \right\}}
\end{align}
where $\textbf{y}_B'=\textbf{W}_B^{-1/2}\textbf{y}_B$, $\textbf{y}_E'=\textbf{W}_E^{-1/2}\textbf{y}_E$,  $f_{b,i,j}=\|\sqrt{\beta P}$ $\textbf{W}_B^{-1/2}\textbf{H}_B
\textbf{d}_{ij}+\textbf{n}_B'\|^2$ and $f_{e,m,k}=\|\sqrt{\beta P}\textbf{W}_E^{-1/2}\textbf{H}_E
\textbf{d}_{mk}+\textbf{n}_E'\|^2$, $\textbf{n}_B'=\textbf{W}_B^{-1/2}(\sqrt{(1-\beta)P} \textbf{H}_B\textbf{T}_{AN}\textbf{n}+ \textbf{n}_B )$, and $\textbf{n}_E'=\textbf{W}_E^{-1/2}(\sqrt{(1-\beta)P} \textbf{H}_E\textbf{T}_{AN}\textbf{n}+ \textbf{n}_E )$, where $\textbf{d}_{ij}=\textbf{x}_i-
\textbf{x}_j$, $\textbf{d}_{mk}=\textbf{x}_m-
\textbf{x}_k$, $\textbf{x}_i,\textbf{x}_j,\textbf{x}_m,\textbf{x}_k$ is one of possible transmit vectors in the set of combining antenna and all possible symbols. Here,  $\textbf{W}_B$ and $\textbf{W}_E$ are the covariance matrices of interference plus noise of Bob and Eve, respectively, where $\textbf{W}_B=(1-\beta)P\textbf{C}_B + \sigma_B^2\textbf{I}_{N_r}$ and $\textbf{W}_E=(1-\beta)P\textbf{C}_E + \sigma_E^2\textbf{I}_{N_e}$, with
$\textbf{C}_B=\textbf{H}_B\textbf{T}_{AN}\textbf{T}_{AN}^H\textbf{H}_B^H$ and
$\textbf{C}_E=\textbf{H}_E\textbf{T}_{AN}\textbf{T}_{AN}^H\textbf{H}_E^H$, respectively. According to \cite{Wang2015Secrecy},  pre-multiplying $\textbf{y}_B$ and $\textbf{y}_E$ by $\textbf{W}_B^{-1/2}$ and $\textbf{W}_E^{-1/2}$ is to whiten colored noise plus AN into an white noise, and  doesn't change the mutual information. In other words,  $I(\textbf{x};\textbf{y}_E)=I(\textbf{x};\textbf{y}_E')$. Finally, the average SR is given as
\begin{align}\label{SR}
\bar{R}_s=\mathbb{E}_{\textbf{H}_B,\textbf{H}_E}\left[ I(\textbf{x};\textbf{y}_B)- I(\textbf{x};\textbf{y}_E) \right]^+.
\end{align}
where $\left[ a \right]^+$=max(a,0) and $R_s =I(\textbf{x};\textbf{y}_B)- I(\textbf{x};\textbf{y}_E)$ is the instantaneous SR for a specific channel realization. Here, we assume that the ideal knowledge of $\textbf{H}_B$ and $\textbf{H}_E$ are available at the transmitter per channel use, which would be true that the eavesdropper is a participating user in a wiretap network \cite{Survey2014}. The optimization problem can be casted as
\begin{align}
\max ~~  R_s~~~~\textrm{subject to} ~~ 0< \beta <1.   \label{P0}
\end{align}

\section{Two Proposed PA Strategies}

\subsection{Proposed GD method}
Due to the expression of SR lacks closed-form, it is hard for us to design a valid method to optimize PA factor effectively. Although exhaustive search (ES) method \cite{Wu2015Secret} can be employed to search out the optimal PA factor, the high complexity restricts its application for secure SM systems. With that in mind, the cut-off rate with closed-form for traditional MIMO systems in \cite{Aghdam2017Joint} can be extended to the secure SM systems, which is an efficient metric to optimize the PA factor below, and  rewritten by
\begin{align} \label{R_s_appro}
R_s^{a}=I_0^{B}-I_0^{E},
\end{align}
where $I_0^{B}$ is the cut-off rate for Bob, and it can be derived by using the following formula
\begin{align} \label{Ib}
I_0^{B}= -\textrm{log}_2\sum \limits_{i=1}^{N_tM}\sum \limits_{j=1}^{N_tM}\frac{1}{(N_tM)^2}\int p(\textbf{y}|\textbf{x}_i)^{1/2} \cdot p(\textbf{y}|\textbf{x}_j)^{1/2} d\textbf{y},
\end{align}
which is viewed as a valid lower-bound on the $I_B$. For a given $\textbf{H}_B$, the receive signal $\textbf{y}_B$ is a complex Gaussian distribution, and the corresponding conditional probability is
\begin{align} \label{pB}
p(\textbf{y}_B|\textbf{x}_i)=\frac{1}{(\pi\sigma_B^2)^{N_r}}\textrm{exp}\left( \|(\textbf{y}'_B-\sqrt{\beta P}\textbf{H}'_B\textbf{x}_i)\|^2 \right),
\end{align}
where $\textbf{y}'_B=\textbf{W}_B^{-1/2}\textbf{y}_B$ and $\textbf{H}'_B=\textbf{W}_B^{-1/2}\textbf{H}_B$.
Making use of (\ref{pB}), we have
\begin{align} \nonumber
I_0^{B}=&2\textrm{log}_2N_tM- \\
&\textrm{log}_2\sum \limits_{i=1}^{N_tM}\sum \limits_{j=1}^{N_tM}\textrm{exp}\left( \frac{-\beta P\textbf{d}_{ij}^H\textbf{H}_B^H\textbf{W}_B^{-1}\textbf{H}_B\textbf{d}_{ij}}{4} \right),
\end{align}
which can be derived  similarly to Appendix A in \cite{Aghdam2017Joint} with a slight modification. Similarly, the cut-off rate $I_0^E$ is given by
\begin{align} \nonumber
I_0^{E}=&2\textrm{log}_2N_tM- \\
&\textrm{log}_2\sum \limits_{m=1}^{N_tM}\sum \limits_{k=1}^{N_tM}\textrm{exp}\left( \frac{-\beta P\textbf{d}_{mk}^H\textbf{H}_E^H\textbf{W}_E^{-1}\textbf{H}_E\textbf{d}_{mk}}{4} \right).
\end{align}
 Since the approximated SR with a closed-form expression is obtained, the optimization problem can be converted onto
\begin{align}
\max ~~  R_s^{a}~~~~\textrm{subject to}~~  0\leq \beta \leq1. \label{P1}
\end{align}
where $R_s^{a}= \textrm{log}_2\kappa_E(\beta)- \textrm{log}_2\kappa_B(\beta)$,
\begin{align}
&\kappa_B(\beta)=\sum \limits_{i=1}^{N_tM}\sum \limits_{j=1}^{N_tM}\textrm{exp}\left( \frac{-\beta P\textbf{d}_{ij}^H\textbf{H}_B^H\boldsymbol{\omega}_B(\beta)\textbf{H}_B\textbf{d}_{ij}}{4} \right),  \\
&\kappa_E(\beta)= \sum \limits_{m=1}^{N_tM}\sum \limits_{k=1}^{N_tM}\textrm{exp}\left( \frac{-\beta P\textbf{d}_{mk}^H\textbf{H}_E^H\boldsymbol{\omega}_E(\beta)\textbf{H}_E\textbf{d}_{mk}}{4} \right),
\end{align}
and $\boldsymbol{\omega}_B(\beta)=\textbf{W}_B^{-1}$, $\boldsymbol{\omega}_E(\beta)=\textbf{W}_E^{-1}$. It is seen that the optimization problem (\ref{P1}) is non-convex because the terms $\textrm{log}_2\kappa_B(\beta)$ and $\textrm{log}_2\kappa_E(\beta)$ of objective function are non-concave. To maximize $R_s^a$, the GD method can be employed to directly optimize the PA optimization variable, and yields the gradient vector of $R_s^a$
\begin{equation}
\begin{aligned}
&\nabla_\beta R_s^{a}=\\
&\frac{P}{\textrm{ln}2\cdot\kappa_B}\sum \limits_{i=1}^{N_tM}\sum \limits_{j=1}^{N_tM} \chi_B \cdot \textrm{exp}\left( \frac{-\beta P\textbf{d}_{ij}^H\textbf{H}_B^H\boldsymbol{\omega}_B(\beta)\textbf{H}_B\textbf{d}_{ij}}{4} \right) \\
&- \frac{P}{\textrm{ln}2\cdot\kappa_E}\sum \limits_{m=1}^{N_tM}\sum \limits_{k=1}^{N_tM} \chi_E \cdot \textrm{exp} \left( \frac{-\beta P\textbf{d}_{mk}^H\textbf{H}_E^H \boldsymbol{\omega}(\beta) \textbf{H}_E\textbf{d}_{mk}}{4} \right) \label{gradient}
\end{aligned}
\end{equation}
where
\begin{align} \label{X_B} \nonumber
\chi_B=0.25\{&\textbf{d}_{ij}^H\textbf{H}_B^H\boldsymbol{\omega}(\beta)\textbf{H}_B\textbf{d}_{ij}+\\
&\beta P\textbf{d}_{ij}^H\textbf{H}_B^H\boldsymbol{\omega}(\beta)\textbf{C}_B\omega(\beta)\textbf{H}_B\textbf{d}_{ij}\},
\end{align}
\begin{align} \label{X_E}  \nonumber
\chi_E=0.25\{&\textbf{d}_{mk}^H\textbf{H}_E^H\boldsymbol{\omega}(\beta)\textbf{H}_E\textbf{d}_{mk}+\\
&\beta P\textbf{d}_{mk}^H\textbf{H}_E^H\boldsymbol{\omega}(\beta)\textbf{C}_E\omega(\beta)\textbf{H}_E\textbf{d}_{mk}\}
\end{align}
where the second term of the right-hand side in (\ref{X_B}) and (\ref{X_E}) holds due to the fact that $\nabla(\textbf{X}^{-1})=-\textbf{X}^{-1}\nabla(\textbf{X})\textbf{X}^{-1}$, where $\nabla(\cdot)$ denotes the gradient operation. So as to get a better PA factor, we can repeat the algorithm with different initial values and find out the best $\beta$ that have the best value of SR. However, it is  guaranteed that the best solution of GD method converges to the global optimal solution as  the number of initial randomizations tends to large-scale.

\subsection{Proposed Max-P-SINR-ANSNR method}
To address the iterative convergence problem of GD, a closed-form solution is preferred.  Now,  AN is viewed as the useful signal of Eve. The SINR at Eve is defined as AN-to-signal-noise ratio (ANSNR). If the product of SINR at Bob and ANSNR at Eve is maximized, it is guaranteed that  at least one of SINR at Bob and ANSNR at Eve or both is high.  This will  accordingly improve SR.  From the definition of SINR,  the SINR of Bob and ANSNR of Eve are defined as
\begin{equation}
\textrm{SINR}_B(\beta)=\frac{\beta Ptr(\textbf{H}_B\textbf{H}_B^H)}{N_t(1-\beta) Ptr(\textbf{C}_B)+N_tN_r\sigma_B^2},
\end{equation}
and
\begin{equation}
\textrm{ANSNR}_E(\beta)=\frac{N_t(1-\beta)Ptr(\textbf{C}_E)}{N_t\beta Ptr(\textbf{H}_E\textbf{H}_E^H)+N_tN_e\sigma_E^2},
\end{equation}
respectively. Observing the above two definitions,  as $\beta$ varies from 0 to 1, $\textrm{SINR}_B$ increases and $\textrm{ANSNR}_E$ decreases. Thus,  they are two conflicting cost functions. If we multiply $\textrm{SINR}_B$ and $\textrm{ANSNR}_E$, their product will form a maximum value at some point in the interval [0, 1] due to their rational property. Their product is defined as follows
\begin{align}
f(\beta)=\textrm{SINR}_B\cdot\textrm{ANSNR}_E=\frac{a_ba_e\beta(1-\beta)}{[(1-\beta)b_b+c_b](\beta b_e+c_e)}
\end{align}
where $a_b=\frac{1}{N_t}Ptr(\textbf{H}_B\textbf{H}_B^H)$, $a_e=Ptr(\textbf{C}_E)$, $b_b=Ptr(\textbf{C}_B)$, $c_b=N_r\sigma_B^2$, $b_e=\frac{1}{N_t}Ptr(\textbf{H}_E\textbf{H}_E^H)$, and $c_e=N_e\sigma_E^2$.
Therefore, the corresponding optimization problem is established as
\begin{align}  \label{fb}
\max\limits_{\beta} \  f(\beta)~~~~ \textrm{subject to}\  0\leq \beta\leq1.
\end{align}
which gives the derivative of cost function $f(\beta)$
\begin{align}   \label{f,b}
f'(\beta)=\frac{df(\beta)}{d\beta}=\frac{-a_ba_e(a\beta^2+2b\beta-b)}{\{[b_b(1-\beta)+c_b](b_e\beta+c_e)\}^2}=0
\end{align}
where $a=c_bb_e-c_eb_b$, and $b=c_eb_b+c_ec_b$. The above equation generates the  two candidate solutions to (\ref{fb})
\begin{align}
  \beta_1=\frac{-b-\sqrt{b^2+ab}}{a},~~\beta_2=\frac{-b+\sqrt{b^2+ab}}{a}.
\end{align}
 Considering the constraint $0\leq\beta\leq1$ of (\ref{fb}),  we have the set of all four potential solutions as follows
 \begin{align}
  S=\left\{\beta_1,~\beta_2,~\beta_3=0,~\beta_4=1\right\}.
\end{align}
 It is clear that $\beta_3=0$ means that no power is allocated to the useful signal, namely no mutual information is sent. In other words, SR equals zero. Since $a$ is positive, we can infer $\beta_1<0$. It is impossible because $\beta$ belongs to the interval [0,~1]. $\beta_1$ and $\beta_3$ should be removed from set $S$. Since the denominator of (\ref{f,b}) is positive, it is clear that $f'(\beta)$ is negative when $\beta>\beta_2$ and $f'(\beta)$ is positive when $\beta<\beta_2$. Therefore, $\beta_2$ is a local maximum point and $\beta_4$ is a local minimum point. Finally, we conclude that the optimal solution to (\ref{fb}) is
\begin{align}
\beta_2=\frac{-b+\sqrt{b^2+ab}}{a}.
\end{align}

\section{Simulation and Numerical Results}
To evaluate the SR performance of the two proposed PA strategies, system parameters are set as follows:  $N_t=4$, $N_r=2$,  $N_e=2$, and quadrature phase shift keying (QPSK) modulation. At the same time, for the convenience of simulation, it is assumed that the total transmit power $P=N_t$ and the noise variances are identical, i.e., $\sigma_B^2=\sigma_E^2$.

\begin{figure}[htbp]
\centerline{\includegraphics[width=0.48\textwidth]{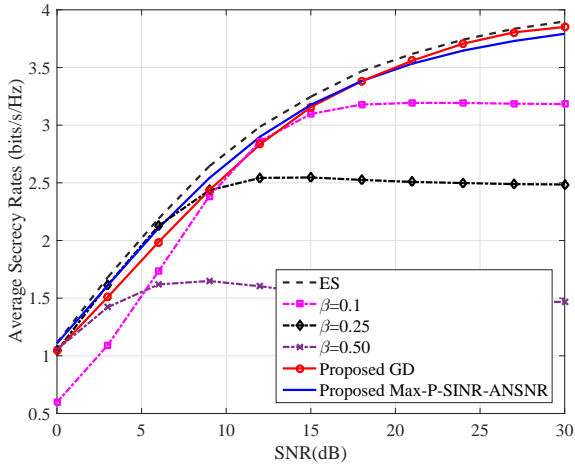}}
\caption{Comparison of average SR versus SNR for different PA strategies with $N_t=4$, $N_r=2$, and $N_e=2$.}
\label{fig1}
\end{figure}

Fig.~\ref{fig1} demonstrates the average SR versus SNR for  different PA strategies, where optimal ES method  is used  as a performance upper bound. It can be clearly seen from Fig.~\ref{fig1} that the performance of the proposed GB and Max-P-SINR-ANSNR  are  closer to the optimal security performance in the low and medium SNR regions. However, the former is slightly worse than the latter in the high SNR region. In all SNR regions, the proposed two methods exceeds three fixed PA strategies in terms of SR. This confirms that optimal PA can improve the SR performance.

\begin{figure}[htbp]
\centerline{\includegraphics[width=0.48\textwidth]{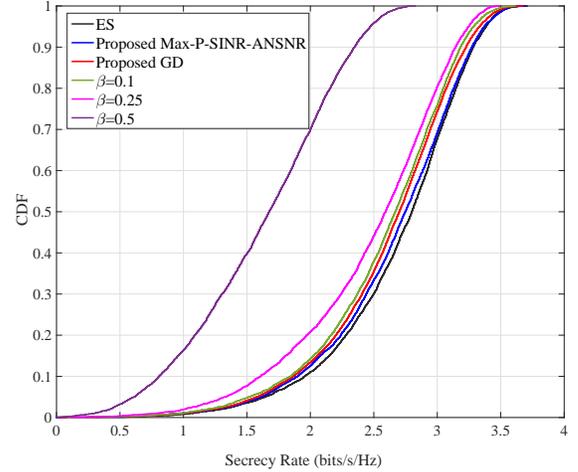}}
\caption{ CDFs of SR  with $N_t=4$, $N_r=2$, $N_e=2$, and SNR=10dB.}
\label{fig2}
\end{figure}

 Fig.~\ref{fig2} plots the cumulative distribution function (CDF)  of  SR for different PA strategies with SNR=10dB. It can be  seen that the CDF curves of the proposed Max-P-SINR-ANSNR and GD are up to the right of those of three fixed PAs. This means that they perform better than three fixed PA strategies. Therefore, the proposed two PA methods have substantial SR performance gains over fixed PAs.

\section{Conclusion}
In this paper, we have made an investigation on  PA strategies for the secure SM system. Here, we proposed two PA strategies: GD and Max-P-SINR-ANSNR. The former is iterative and the latter is closed-form. In other words, the latter is of low-complexity. Simulation results showed that the proposed GD and Max-P-SINR-ANSNR methods  nearly achieve the optimal SR performance achieved by ES. The former is better than the latter in the high SNR region, and worse than the latter in the low to medium regions in terms of SR performance.

%

\vspace{12pt}

\bibliographystyle{IEEEtran}
\bibliography{IEEEabrv,ref}

\end{document}